\documentclass[letters,fleqn,usenatbib,useAMS]{mnras}

\usepackage{graphicx}
\usepackage{amssymb}
\usepackage{url}
\usepackage[dvipsnames]{xcolor}

\def\TheEvent{GW150914}
\def\neventstwoifos{237}
\def\neventsthreeifos{195}
\newcommand{\si}{\ensuremath{\sim}}

\def\linf{\texttt{lalinference}}
\def\gstlal{\texttt{gstlal}}
\def\cwb{\texttt{cWB}}
\def\lib{\texttt{LIB}}
\def\bayestar{\texttt{BAYESTAR}}

\def\degtwo{deg$^2$}
\def\mmintwoifos{20$M_\odot$}

\renewcommand{\cite}{\citep}

\newcommand{\sv}[1]{\textcolor{black}{#1}}

\title{On similarity of binary black hole gravitational-wave skymaps: to observe or to wait?}
\author[Salvatore Vitale et al.]{
Salvatore Vitale$^{1}$\thanks{e-mail: \href{mailto:salvatore.vitale@ligo.org}{salvatore.vitale@ligo.org}}, Reed Essick$^{1}$, Erik Katsavounidis$^{1}$, \newauthor Sergey Klimenko$^{2}$ and Gabriele Vedovato$^{3}$
\\
$^{1}$LIGO, Massachusetts Institute of Technology, Cambridge, Massachusetts 02139, USA\\
$^{2}$University of Florida, P.O.Box 118440, Gainesville, Florida, 32611, USA\\
$^{3}$INFN, Sezione di Padova, via Marzolo 8, 35131 Padova, Italy}
\begin{document}
\label{firstpage}
\pagerange{\pageref{firstpage}--\pageref{lastpage}}
\maketitle

\begin{abstract}
Localization estimates for \TheEvent, the first binary black hole detected by the LIGO instruments, were shared with partner facilities for electromagnetic follow-up. 
While the source was a compact binary coalescence (CBC), it was first identified by algorithms that search for unmodeled signals, which produced the skymaps that directed electromagnetic observations. 
Later on, CBC specific algorithms produced refined versions, which showed significant differences. 
In this paper we show that those differences were not accidental and that CBC and unmodeled skymaps for binary black holes will frequently be different; we thus provide a way to determine whether to observe electromagnetically as promptly as possible (following a gravitational-wave detection), or to wait until CBC skymaps become available, should they not be available in low latency. We also show that, unsurprisingly, CBC algorithms can yield much smaller searched areas.
\end{abstract}

\section{Introduction}

The first-ever gravitational-wave detection, the black-hole binary (BBH) \TheEvent, was identified in low-latency by algorithms that search for unmodeled signals (henceforth, burst algorithms)\citep{2016PhRvD..93l2004A}: \cwb{} \cite{2016PhRvD..93d2004K} and \lib{}~\citep{2015arXiv151105955L}.
These algorithms produced skymaps in low-latency and these were circulated to partner facilities for electromagnetic (EM) and neutrino follow-up with a latency of \si2 days~\cite{2016ApJ...826L..13A,2016PhRvD..93l2010A} (All the GCN pertinent to \TheEvent{} can be found at http://gcn.gsfc.nasa.gov/other/GW150914.gcn3). Skymaps were later generated by a fast sky localization algorithm for compact binary coalescence (CBC) sources called \bayestar~\cite{2016PhRvD..93b4013S,2016ApJ...826L..13A} and by a full parameter estimation algorithm that assumed a CBC model, \linf~\cite{2015PhRvD..91d2003V,GW150914-PARAMESTIM}, and found to be different from the burst results. For example, the \linf{} skymap released on January 13 2016 and had a 90\% confidence level area of 620~\degtwo, less than half of which overlapped with the 90\% area of the burst skymaps~\cite{2016ApJ...826L..13A,2016PhRvD..93l2004A}. Many groups have already reported on their follow-up observations of \TheEvent~\cite{2016ApJ...826L...6C,2016ApJ...820L..36S,2016MNRAS.462.4094S,2016ApJ...823L..33S,2016MNRAS.460L..40E,2016ApJ...823L...2A,2016ApJ...825L...4T,2041-8205-829-1-L12,2016ApJ...829L..28P,2016ApJ...829L..34G,2016ApJ...828L..16D}.

The skymap provided by \linf{} is considered by the LIGO-Virgo collaboration to provide the definitive sky localization for this (and every) CBC source~\cite{2016ApJ...826L..13A} and for this reason we will use such maps for the purpose of the study. The reconstruction of these two burst algorithms together with the \linf{} CBC skymap was reported in~\citet{2016PhRvD..93l2004A}\footnote{FITS files for all skymaps can be found at \citet{LOSC-GW150914}.}. 
Two types of differences are visible: from one side different parts of \emph{the same} sky-ring (defined as the locus of points with fixed time-of-arrival difference between the LIGO instruments~\cite{2016LRR....19....1A}) have posterior support; from the other, slightly different rings can have posterior support~\cite{2016ApJ...826L..13A}. 
Until the CBC skymaps were provided, EM partners observed regions of the sky where only the burst algorithms had support. 

We note that only burst algorithms found \TheEvent{} in low-latency because CBC specific algorithms did not search for BBH at that time~\cite{2016ApJ...826L..13A}, since BBH are not traditionally expected to be luminous in the EM. 
Later in the run, a low-latency BBH version of the CBC search algorithm \gstlal~\cite{2012ApJ...748..136C,2014PhRvD..89b4003P} was enabled, which can be followed-up by the low-latency CBC sky localization algorithm \bayestar~\cite{2016PhRvD..93b4013S}.
It is thus plausible that from the second observing run onward only CBC skymaps will be released for BBH events, with \bayestar{} estimates available in close to real time~\cite{2016PhRvD..93b4013S} and with the definitive \linf{} maps obtained typically within a few days~\cite{2015PhRvD..91d2003V}.
However, one may envision situations where skymaps obtained by the low-latency burst searches become available prior to the CBC ones. This can happen if the astrophysical signal deviates from the waveform template used by the CBC search algorithms. For example, spin-induced amplitude and phase modulation might be present in the signal, which is not captured by the spin-aligned waveforms used by CBC searches~\cite{GW150914-CBC,O1BBH}, causing a loss of signal-to-noise ratio (SNR)~\cite{GW150914-CBC}. 

Efforts to introduce fully spinning CBC searches are ongoing, but not yet implemented~\cite{2016PhRvD..94b4012H}.
Given their unmodeled nature, burst algorithms would not be affected in the same way. In this example, CBC skymaps could be produced with higher latency, after fully precessing spin parameter estimation analyses are completed.
Situations in which the burst skymaps are generated earlier could also be produced by technical issues in processing the data or race conditions that lead to latency offsets. Additionally, because of the different strides used by each of the burst and CBC low latency searches, it is plausible that their (temporal) overlaps may not be perfect~\cite{2016PhRvD..93l2004A,GW150914-CBC}.
Although we do not expect these situations to be common, they certainly warrant consideration.

Potential differences in the time at which skymaps may become available can have significant implications for EM observing strategies designed to maximize the chance of capturing an EM counterpart. It is thus important to verify if the degree of difference between CBC- and burst-reconstructed BBH skymaps similar to GW150914 would be typical and, overall, quantify the relation of such skymaps.
In this paper we consider a large set of BBH simulations and compare their skymaps produced by the most recent version of the low-latency burst algorithms that ran in the first observing run (O1), \cwb{} and \lib{} \cite{2016PhRvD..93d2004K,2015arXiv151105955L} and by the CBC parameter estimation algorithm \linf~\cite{2015PhRvD..91d2003V}.

\section{Simulations}

The sky localization performance of burst algorithms was thoroughly analyzed in \citet{2015ApJ...800...81E}, where several morphologies traditionally used to simulate GW bursts were investigated, including BBH signals generated by heavy black holes. 
\citet{2015ApJ...800...81E} focuses on two burst algorithms, considering a  two-interferometer (Hanford, WA and Livingston, LA, denoted as ``HL'') network with the two LIGOs at early 2015 sensitivity (roughly achieved during O1), and a three-interferometer (HLV, where ``V'' denotes Virgo) network with the two LIGOs and Virgo at their early sensitivity~\citep{2016LRR....19....1A}. 
In this paper we extend that study to include skymaps from the CBC parameter estimation algorithm \linf{}~\cite{2015PhRvD..91d2003V}.
Of the original BBH events considered in \citet{2015ApJ...800...81E} we only kept sources with component masses larger than \mmintwoifos, because heavy BBH are more compact in the time-frequency domain and thus more easily detectable by burst algorithms with high confidence (For example, the second BBH detection, GW151226, was not found with high significance by burst algorithms~\cite{GW151226-DETECTION}).
This selection leaves \neventstwoifos{} simulated events for the HL network and \neventsthreeifos{} simulated events for the HLV network.

We still consider the HL network because the first few months of the second observing run (O2) will only comprise the two LIGOs, with a sensitivity not too different from O1. We therefore expect the HL runs of~\citet{2015ApJ...800...81E} to be representative of the events one could detect in the first part of O2, without Virgo. Similarly, the HLV events could represent the typical situation for the late part of O2, with Virgo online.
The simulated events were added to simulated Gaussian noise (identical to \citet{2015ApJ...800...81E}) and their parameters estimated with the nested sampling version of \linf~\cite{2015PhRvD..91d2003V}. \citet{2015ApJ...804..114B} showed that using Gaussian noise yields the same statistical distribution of skymaps one would obtain with real interferometric noise.

We used a spin-aligned template family (\texttt{SEOBNRv2ROM}~\cite{2014CQGra..31s5010P,2016PhRvD..93f4041P}) to decrease the computational time.
The simulated signals were also spin-aligned, although from a different waveform family (\texttt{IMRphenomB}~\cite{2011PhRvL.106x1101A}). 
Because the typical detection with advanced detectors will have a small inclination angle~\cite{2011CQGra..28l5023S}, as indeed was the case for both \TheEvent{} and GW151226~\cite{GW150914-PARAMESTIM,O1BBH,GW151226-DETECTION}, which reduces the visible effects of eventual spin precession~\cite{PhysRevD.49.6274,PhysRevLett.112.251101}. This spin-aligned study thus represents a reasonable approximation. Furthermore, it has been shown that neglecting mild spin precession does not significantly affect the reconstructed skymaps~\cite{2016ApJ...825..116F,2009CQGra..26k4007R}.
We allowed for calibration uncertainties in the data, using the same method described in \cite{GW150914-PARAMESTIM}. 
For the amplitude calibration error prior, we used a zero-mean Gaussian with standard deviation of 10\%, while for the phase we used a standard deviation of 5 degrees. This is comparable with what was achieved during O1~\cite{GW150914-CALIBRATION,O1BBH}.

The burst algorithms skymaps originally released in \citet{2015ApJ...800...81E} were regenerated using the most recent versions of \cwb{} and \lib.  The burst algorithms do not currently marginalize over calibration uncertainties.

\section{Results}

\subsection{Fidelity}

\begin{figure}
    \includegraphics[width=0.95\linewidth]{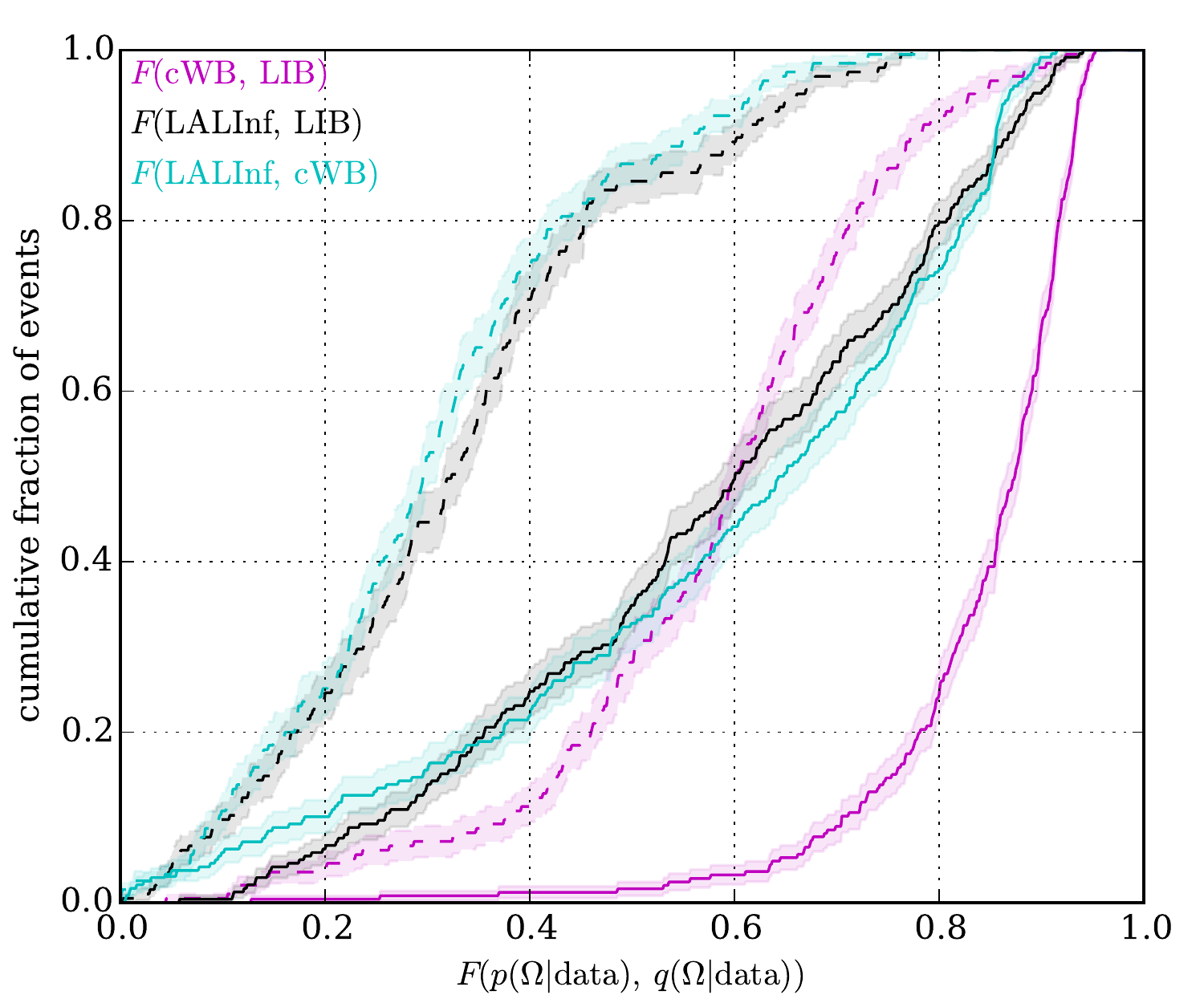}
    \caption{The cumulative distribution of the Fidelity of skymaps for pairs of algorithms in the HL (solid) and HLV (dashed) analyses.} 
    \label{Fig.Fidelity}
\end{figure}

We first address the question of how similar two skymaps are. 
Several figures of merit can be used to quantify similarity. 
In this section we use the Fidelity, defined as follows. Given two skymaps $p$ and $q$ we calculate the Fidelity as:

\begin{equation}\label{e:fidelity}
    F(p,q)\equiv \sum_{i}{\sqrt{p_i q_i} }  \in [0,1]
\end{equation}

\noindent where $p_i$ and $q_i$ are the probabilities assigned to the $i^\mathrm{th}$ pixel by each skymap. 
The Fidelity would be exactly 1 for two identical skymaps, whereas it would be 0 for skymaps which do not have any common support. 
In Fig.~\ref{Fig.Fidelity} we report the cumulative distribution of the Fidelity for pairs of algorithms for the HL (solid lines) and HLV (dashed lines) networks.

Let us consider the HL network first. We see that the two burst algorithms generally agree very well with each other, with 50\% of skymaps having a Fidelity larger than $0.85$. 
The \cwb-\lib{} Fidelity for \TheEvent{} (0.55~\cite{2016PhRvD..93l2004A}) was significantly lower than this; which we explain later.  
Both burst algorithms have a similar degree of Fidelity with \linf. For 50\% of the events, the Fidelity of \cwb{} skymaps with \linf{} skymaps is larger than \sv{$0.65$}, while for \lib{} this number is \sv{$0.6$}. 
These numbers are consistent with what found for \TheEvent{} (0.51 for \cwb-\linf{} and 0.68 for \lib-\linf).

\begin{figure}
    \includegraphics[width=0.5\textwidth]{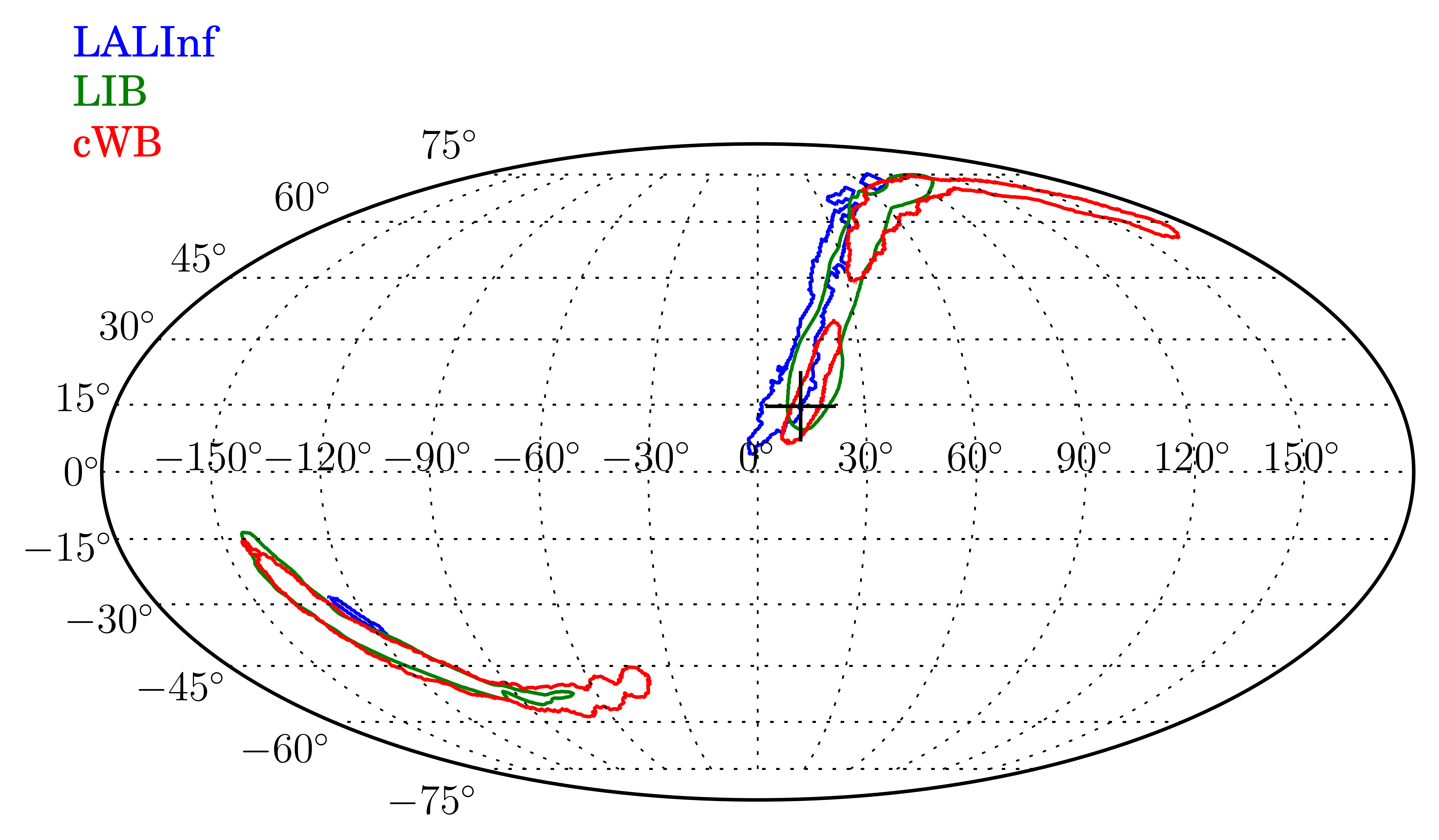}
    \caption{HLV event 968666271. Burst and CBC pick a different ring. The true position is indicated with a black ``+''. The Fidelity of the burst skymaps with \linf{} is below 0.35.}
    \label{Fig.DiffRing2016}
\end{figure}

Let us now consider the HLV network, which might realistically be online in early 2017~\cite{2016LRR....19....1A}. 
These are the dashed lines in Fig.~\ref{Fig.Fidelity}. 
Comparing with the full lines we see a general degradation of the Fidelity. 
In fact, the median Fidelity across burst algorithms reduces to roughly 0.65, while it is above 0.8 for the HL network. 
The degradation is even more significant for the Fidelity of each burst algorithm with \linf, with medians at around 0.3. 
This is because the HLV skymaps are known to break up into several islands of probability, for both CBC \cite{2014ApJ...795..105S} and burst algorithms \cite{2015ApJ...800...81E}.
Because \linf{} can better match CBC signals in the data, it makes better use of information from Virgo and fragments first. 
This produces the significant differences between burst and \linf{} localizations with three detectors.
As mentioned earlier, the skymaps from two classes of algorithms might be in tension because they select different parts of the same sky-ring, or because they prefer different rings. 
An example of a source for which different skymaps pick different rings is given in Fig.~\ref{Fig.DiffRing2016}.

\begin{figure}
    \includegraphics[width=0.95\linewidth]{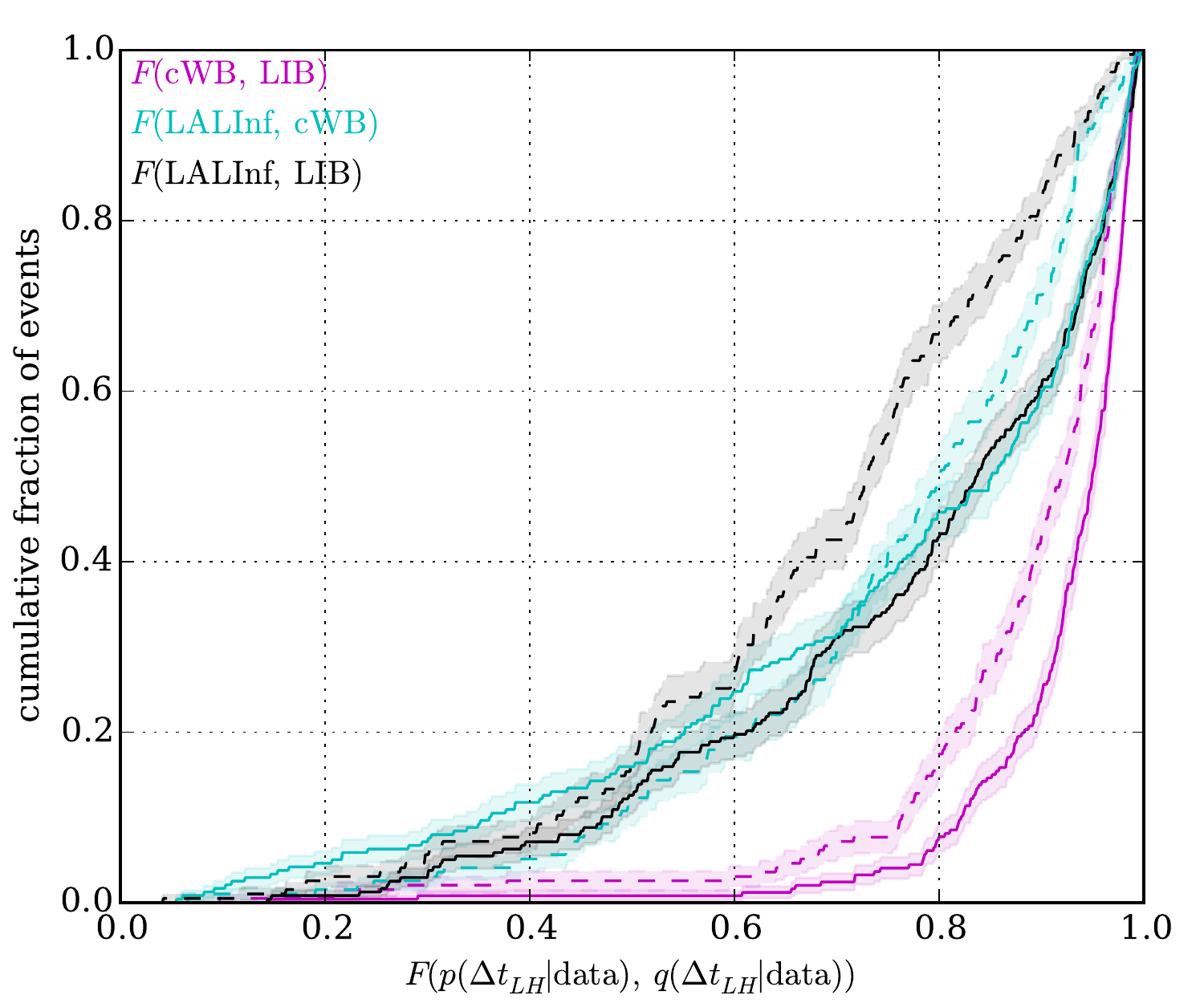}
    \caption{The cumulative distribution of time-of-flight Fidelity for pairs of algorithms in the HL (solid) and HLV (dashed) analysis.} 
    \label{Fig.TOF}
\end{figure}

Can the two effects be disentangled? 
If different rings are selected, one will find that the posteriors on the time-of-flight between two detectors (e.g. Hanford and Livingston) are different. 
To verify how often this happens, we calculate the Fidelity for pairs of one-dimensional posteriors on the HL time-of-flight (Fig.~\ref{Fig.TOF}). 
We see that with the HL network (solid lines), for 50\% of the skymaps the Fidelity is above 0.8 for CBC-burst skymaps, and above 0.9 for \lib-\cwb{} skymaps. The CBC-burst Fidelity is a bit lower for the HLV network, while the \lib-\cwb{} median Fidelity stays close to 0.9. The two burst algorithms thus select the same sky ring very often. 
Strong correlation exists between the Fidelity of the full skymap and that of the time-of-flight, implying that when slightly different rings are selected (low $\Delta t_{HL}$ Fidelity) the overall Fidelity is strongly reduced. 
Thus, the disagreement we found while comparing the full skymaps, Fig.~\ref{Fig.Fidelity}, is determined by both the placement of probability around the ring and the position of the ring itself, with the latter happening less frequently but having a large impact when present.

\begin{figure}
    \includegraphics[width=0.95\linewidth]{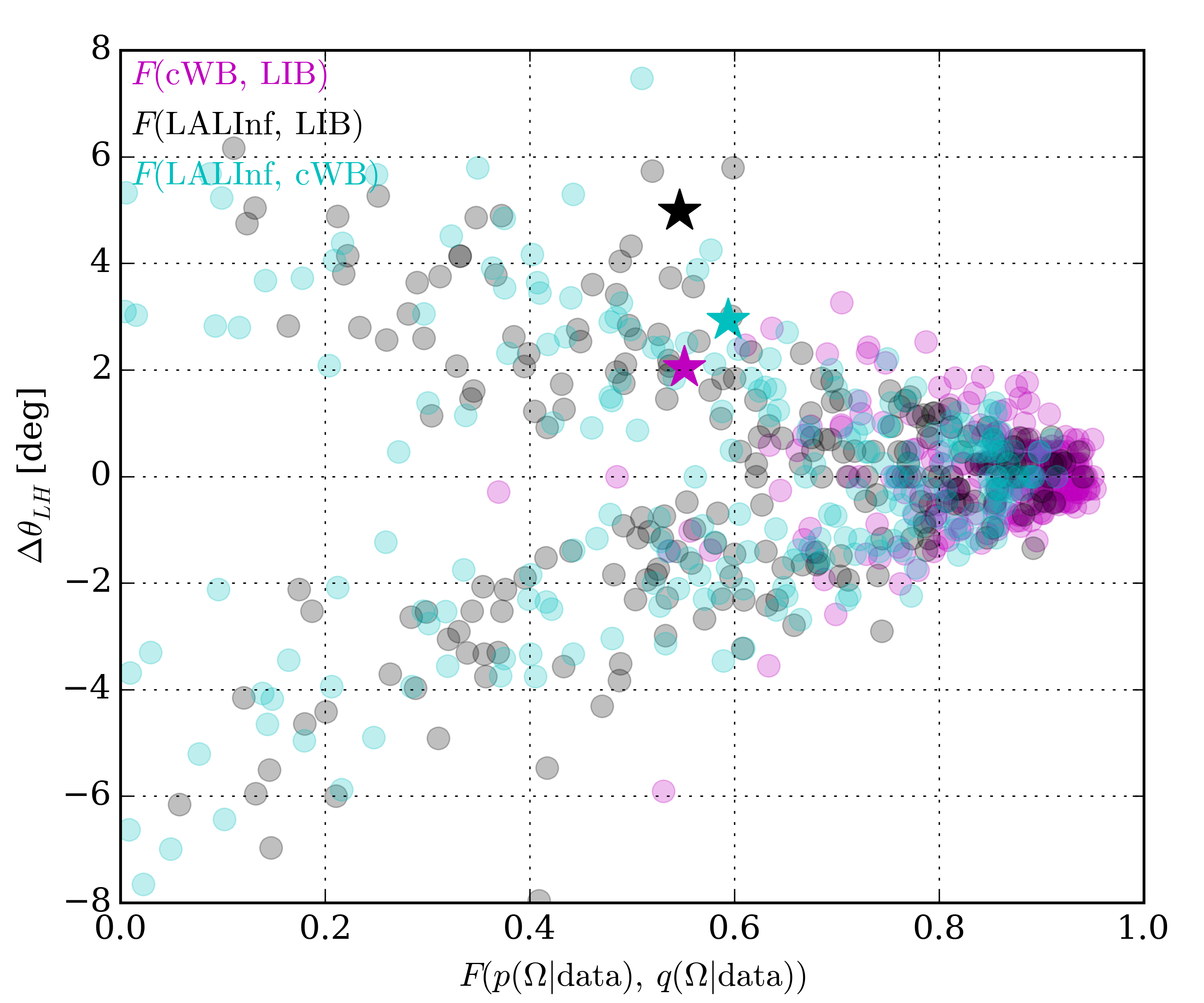}
    \caption{
             A scatter plot of the fidelity using the full skymap and the angular separation between triangulation rings. 
             \TheEvent{}'s values are shown as stars.
            }
    \label{Fig.Scatter}
\end{figure}

Fig.~\ref{Fig.Scatter} shows the clear correlation between angular offsets and decreased Fidelity using data from the HL network, as well as the corresponding values from \TheEvent{} (stars), which fall within the scatter observed in simulations.
We note that while there is quite a bit of scatter, the general trends seem to be independent of which algorithms are actually compared. 
This further suggests that low Fidelity is driven by differences in the selected triangulation rings, because all algorithms produce triangulation rings.

We mentioned above that the \TheEvent{} Fidelity for \lib-\cwb{} was much lower than average. 
We can now understand why: the \lib{} and \cwb{} skymaps for \TheEvent{} have support on different sky rings (see \citet{2016ApJ...826L..13A}). They thus belong to this rarer class of reconstructed skymaps for which there is a significant drop of Fidelity.

Furthermore, one can compute the distribution of the angular distance between rings. 
We find that the distributions look unimodal, with mean smaller than 0.2$^\circ$ for all pairs of algorithms, and standard deviations in the range \sv{$[2.5^\circ-4^\circ]$ ($[2.5^\circ-7^\circ]$) for the HL (HLV) network.}

\subsection{Searched Area}

So far we have focussed on the degree of similarity of skymaps, which was naturally prompted by the fact that burst and CBC skymaps were so different for \TheEvent. 
However that is probably not the most relevant figure of merit to assess the chances of a successful EM follow-up. 
The fact that a skymap is larger or different might not lead to wasted telescope observation time as long as most of the probability is at the true position.

\begin{figure}
    \includegraphics[width=0.95\linewidth]{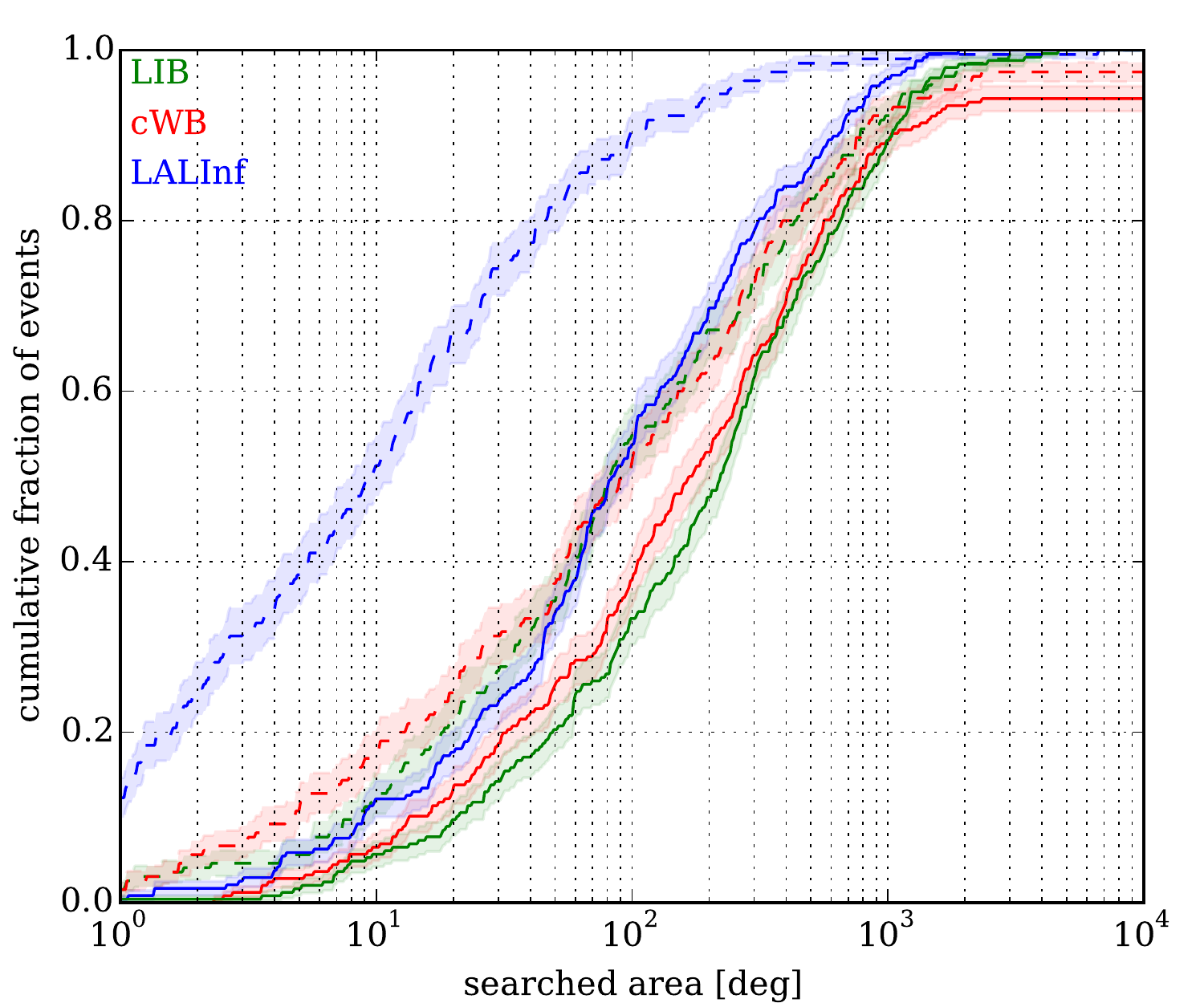}
    \caption{The cumulative distribution of the searched area for each algorithm in the HL (solid) and HLV (dashed) analyses.} 
    \label{Fig.Searched}
\end{figure}

In this section we report the searched area found by the burst and CBC algorithms. 
We define the searched area as the amount of sky one needs to image, starting from the most likely skymap pixel and going down, until the true position is found.
This is shown in Fig.~\ref{Fig.Searched} for HL (solid lines) and HLV (dashed lines). 
Once again, we first focus on the HL network. 
As already seen in \citet{2015ApJ...800...81E}, the two burst algorithms perform similarly, with 50\% of events found with searched areas of roughly $200$ \degtwo. 
We also find that \linf{} can do a factor of 2 better, with 50\% of events having searched areas slightly below $100$ \degtwo. 
This is perhaps not surprising since \linf{} can match the signal in the data nearly exactly\footnote{Nearly exactly but not exactly since we used a different waveform families to simulate and recover the BBH signals.}.
Considering the HLV results, we see that while burst skymaps improve only by a factor of 2, the CBC median searched area reduces by an order of magnitude.

We should note that burst searches can be tuned to specific astrophysical systems, including CBC. Such tuning can yield improved search result, including better localization. In this study we only considered skymaps obtained with the most unconstrained versions of the burst algorithms. We have verified that imposing polarization constraints on \cwb{} can reduce the searched area by a factor of \si 4 for the HLV events.

\section{Discussion and conclusions}

Four skymaps for \TheEvent{}  were sent to partner EM and neutrino facilities: two produced by unmodeled burst algorithms and two by CBC parameter estimation algorithms. 
They showed significant differences and large patches of sky where only one of the skymaps had posterior support.
In this paper, we addressed the question of whether such differences were just accidental or will rather be frequently encountered.

We considered both a two detector (HL, made of the two LIGO instruments) and a three detector (HLV, made of the two LIGO and Virgo) network. 
We used sensitivities comparable to what is expected in the next observing run, starting in the fall of 2016.
Simulating roughly 200 BBH for each network, we compared the skymaps produced by the two low-latency burst analyses that ran during the first observing run, and the skymap produced by the CBC parameter estimation algorithm \linf.
We quantified the degree of similarity between pairs of skymaps using the Fidelity, which is 1 for identical probability distributions and 0 when no overlap exists.
We found that the two burst algorithms typically agree with each other, with median Fidelity of \sv{\si0.85} for the HL network and \sv{\si0.6} for the HLV network. The skymaps are less similar for the HLV network.

This is due to the different way each algorithm deals with sub-threshold events in Virgo, which is less sensitive than either LIGO. The agreement between each of the burst algorithms and the CBC result is lower, around 0.6 for HL and 0.3 for HLV. 
We then verified that the differences can arise from a combination of two factors: different parts of the same sky rings can be selected by different algorithms, or different sky rings can be preferred in the first place. 
This latter effect is less common for HL networks, but significantly decreases the Fidelity. 
We find that the mean angular distance between rings is below \sv{0.2} degrees for all pairs of algorithms, with standard deviations of \sv{~\si 2$^\circ$ to 7$^\circ$.}
Finally, we focused on the searched area, i.e. the amount of sky that must be imaged before the true position of the source is found. 

We found that for the HL network the two burst algorithms lead to similar results, with median searched area of \si200 \degtwo.  The CBC algorithm does a factor of \si 2 better. When the HLV network is considered, we found that using CBC skymaps can lead to a dramatic reduction in searched area. 
In fact, while the median searched area shrinks by only a factor of \si2 for the burst skymaps, it reduces by an order of magnitude for \linf{}, for which the median searched area is only 10~\degtwo{} with an HLV network.
We underline that \linf{} will not produce skymaps with minute-scale latency. 
However, it has been shown that the low-latency algorithm \bayestar{} produces skymaps which agree well with the \linf{} ones~\cite{2014ApJ...795..105S}, at least for the HL network. 
Nonetheless, we used the \linf{} skymaps as representative of the definitive skymaps for a BBH source, which was the LIGO and Virgo collaborations' approach in O1~\cite{2016ApJ...826L..13A,GW150914-PARAMESTIM}.
Furthermore, while \bayestar{} needs the output of a CBC search algorithm to run~\cite{2016ApJ...826L..13A}, \linf{} does not, and can thus refine skymaps for events found by either burst or CBC search algorithms.

Our study shows that the level of tension observed between burst and CBC skymaps for \TheEvent{} is not exceptional (although the level of disagreement of \lib{} and \cwb{} was larger for \TheEvent{} than on average for BBH).
Furthermore, we conclude that not only will low-latency burst and CBC skymaps be different, but the latter will be significantly more accurate.
Waiting for skymaps from CBC analysis, if not available in low latency, may delay EM observations. 
These delays should be considered depending on the EM facility and the possible astrophysical source. 
Commencing observations promptly is the only way to capture an EM signature on timescales less than a day, typically expected for the high energy emission and UV/O/IR~\cite{2012ApJ...746...48M}. 
Alas, this may add uncertainty and hinder success due to the limitations of prompt localization. 
On the other hand, EM facilities targeting emission on the scale of days to months to years (primarily in the radio~\cite{2012ApJ...746...48M}) may opt to wait for subsequent communications from the LIGO-Virgo collaborations regarding localization information.

\section{Acknowledgments}

The authors would like to thank C.~Berry, T.~Dent, V.~Kalogera and S.~Eikenberry, as well as the MNRAS referee for useful comments and suggestions.
The authors acknowledge the support of the National Science Foundation and the LIGO Laboratory. LIGO was constructed by the California Institute of Technology and Massachusetts Institute of Technology with funding from the National Science Foundation and operates under cooperative agreement PHY-0757058.
S.~K thanks the National Science Foundation for support under grant PHY-1505308.
The authors would like to thank the Albert Einstein Institute in Hannover, supported by the Max-Planck-Gesellschaft, for use of the Atlas high-performance computing cluster.  This is LIGO Document P1600300

\bibliographystyle{mnras}
\bibliography{draft.bib,pe.bib}

\end{document}